\documentclass[preprint, showpacs,amsmath,amssymb]{revtex4}
\usepackage{graphicx}
\usepackage{bm}
\begin{document}
\title{Competition between phonon superconductivity and Kondo screening 
in mixed valence and heavy fermion compounds.}
\author{Victor Barzykin} 
\affiliation{Department of Physics and Astronomy, University of Tennessee,
Knoxville, TN  37996-1200}
\author{L. P. Gor'kov}
\altaffiliation[Also at ]{L.D. Landau Institute for Theoretical Physics,
Chernogolovka, 142432, Russia}
\affiliation{National High Magnetic Field Laboratory,
Florida State University,
1800 E. Paul Dirac Dr., Tallahassee, Florida 32310 }
\begin{abstract}
We consider
competition of Kondo effect and
s-wave superconductivity
in heavy fermion and mixed valence superconductors, using the  phenomenological 
approach for the periodic Anderson
model. Similar to the well known results for single-impurity
Kondo effect in superconductors, we have found principal possibility of a re-entrant
regime of the superconducting transition temperature, $T_c$,
in heavy fermion superconductors in a narrow range
of model parameters and concentration of f-electrons. Suppression of $T_c$ in
mixed valence superconductors is much weaker.
Our theory has most validity in the low-temperature 
Fermi liquid regime, without re-entrant behavior of $T_c$. To check its applicability, we performed the fit 
for the $x$-dependence of $T_c$ in Ce$_{1-x}$La$_x$Ru$_3$Si$_2$ and 
obtained an excellent agreement with the experimental data, although no re-entrance
was found in this case. Other experimental
data are discussed in the light of our theoretical analysis. In particular,
we compare temperatures of the superconducting transition for some known homologs,
i.e., the analog periodic lattice compounds with and without f-elements. For a few
pairs of homologs superconductivity exists only in the heavy fermion materials,
thus confirming uniqueness of superconductivity mechanisms for the latter. We 
suggest that for some other compounds the value of $T_c$ may remain of the same
order in the two homologs, if superconductivity originates mainly on some light
Fermi surface, but induces sizable superconducting gap on another Fermi surface,
for which hybridization or other heavy fermion effects are more significant.
By passing, we cite the old results that show that the jump in the specific heat at the
transition reflects heaviness of carriers on this Fermi surface independently of mechanisms
responsible for superconductivity.
\end{abstract}
\vspace{0.15cm}

\maketitle

\section{Introduction}

Since the discovery of the first heavy fermion (HF) system, CeAl$_3$, in 1975 \cite{andres}, these compounds  have attracted 
enormous theoretical and experimental interest due to their fascinating properties. 
The most striking feature of these systems, their extremely large effective mass $m^* \sim (100 - 1000) m_e$ 
of charge carriers, is qualitatively understood in terms of unification of magnetic degrees of freedom with the ones
of itinerant electrons. The magnetic and superconducting properties of
these compounds are both rich and puzzling (For a review, see \cite{hewson,coleman}).  
Unconventional superconducting properties  may be independent of mechanism, and related to the unusual symmetry of the order parameter  
(for reviews, see Refs. \cite{gorkov, SU}). However, unlike common superconductors, the microscopic mechanisms of 
superconductivity (SC) in these materials for the most part remain  unknown. At least a partial answer to it could
be in the question whether conventional, i.e., phonon-mediated superconductivity\cite{BCS} is excluded for them. 
In what follows, we address this issue.

The theoretical framework for studying heavy fermion metals is the 
periodic Anderson model, considered below in section II. Section III sums up the results of the
phenomenological approach. In sections IV and V the main theoretical formulas for s-wave superconductivity
are derived in the same frameworks. In section VI we discuss some scarce experimental data to 
show that, although there are cases where new mechanisms seem to be necessary, in many other cases phonon-mediated
exchange may remain as the cause of superconductivity. In section VII we draw our conclusions. 

The single site Anderson impurity problem in a normal metal has been solved\cite{hewson,tsvelick}. 
However, even for the low impurity concentration,
the competition of superconductivity and Kondo effect or mixed valency  remains a problem, for which
an exact answer was not obtained theoretically.  It is well known that the pair-breaking action of scattering on 
magnetic impurities in s-wave 
superconductors leads to a drastic suppression of the transition temperature with increasing impurity concentration.
Taking the Kondo screening into account, however, this answer
is less obvious, since at low temperatures the Kondo singlet state acts as a non-magnetic impurity\cite{hewson},
and thus a finite concentration of such impurities may not significantly change $T_c$.
 Competition of superconductivity and Kondo effect in alloys has been studied using various approximate methods, starting
from the pioneering work of M\"{u}ller-Hartmann and Zittartz\cite{MH}, and we summarize the results below briefly.

We stress, however, that our main interest lies in the study of superconductivity of dense systems, especially
of stoichiometric heavy fermion compounds, where at low temperatures the Fermi liquid regime becomes restored,
and theoretical methods based on impurity \textit{scattering} lose ground. We shall try to make use of the fact that many
Ce- and U-based compounds have their homologs, i. e., stoichiometric compounds with non-magnetic elements like Y, La, Lu, 
substituting the rare-earth or actinide elements. For a number of them continuous alloy composition range is 
available to trace whether the superconducting transition temperature varies drastically from a phonon-like $T_c$
for non-magnetic compounds to a new mechanism in HF- or mixed valence (MV) compounds. 

\section{Limiting regimes in concentration and model parameters.}

M\"{u}ller-Hartmann and Zittartz\cite{MH} used Nagaoka decoupling scheme for the Green's functions.
The Nagaoka approximation fails at temperatures below the Kondo temperature, yielding non-analytic
features in all physical properties. Thus, this scheme is expected to fail in the Fermi liquid regime, when the
superconducting transition temperature is much less than the Kondo temperature. Nevertheless,  this
theory\cite{MH} has been successfully applied to many Kondo alloys, such as (La, Ce) Al$_2$, (La,Th)Ce, and other Ce compounds.

  The main result\cite{MH} is that, instead of the usual paramagnetic pair breaking
curve\cite{AG}, the dependence of $T_c$ on concentration acquires, due to the Kondo screening, a characteristic ``S''-shape with 
a re-entrant behavior.
However, $T_c$ never goes to zero at low temperatures. Such re-entrant behavior of $T_c$ has been observed in the
heavy fermion alloys (La,Ce)Al$_2$\cite{riblet,maple}, (La,Th)Ce\cite{huber}, 
and (La,Y)Ce\cite{winzer, dreyer}. In particular,
three transition temperatures were clearly seen in La$_{.7915}$Ce$_{0.0085}$Y$_{.20}$\cite{winzer}.
As the temperature was lowered, the first transition $T_{c1}$ (from normal to superconducting state) was observed
at $0.55$K, the second transition $T_{c2}$ (from superconducting to normal state) at $0.27$K, and the third 
transition (back to superconducting state) at $0.05$K. Significant deviations from the Abrikosov-Gor'kov theory were also
observed in Kondo superconductors LaCe and LaGd\cite{chaikin}, and PbCe and InCe films \cite{delfs}.

  In its simplest form, the single-impurity Anderson
Hamiltonian has the following form:
\begin{eqnarray}
\label{singA}
H_{single} & = & \sum_{\bm{k} \sigma} \epsilon_k c^{\dagger}_{\bm{k} \sigma} c_{\bm{k} \sigma} +
\sum_{\sigma} E_0 f^{\dagger}_{\sigma} f_{\sigma} \\  
& & + V \sum_{\bm{k} \sigma} (c^{\dagger}_{\bm{k} \sigma} f_{\sigma} + H.c.) + U n_{\uparrow} n_{\downarrow} \nonumber,
\end{eqnarray}
where $E_0$ is the energy of a localized orbital, $U$ is the on-site Coulomb repulsion energy, and
$V$ is the hybridization integral between localized states and conduction states $\epsilon_k$. The broadening
of the local level due to hybridization is given by the golden rule:
\begin{equation}
\Gamma = \pi \rho V^2,
\label{gam}
\end{equation}  
where $\rho$ is the single-spin density of states at the Fermi energy. The possible $f$-configurations for $Ce$ and $U$
are $f^0$, $f^1$, and $f^2$. For Yb ($f^{12}$, $f^{13}$, $f^{14}$), one can treat Eq.(\ref{singA}) as the Hamiltonian for holes.

The renormalization group analysis of the model Eq.(\ref{singA}) has been performed by Haldane\cite{haldane}, who has shown
that there are two fixed points:

\noindent
(1) \textbf{The Kondo regime}, 
\begin{equation}
- \epsilon_f \gg \Gamma,
\end{equation}
in which the renormalized impurity level $\epsilon_f$ stops well below the chemical potential and at very large $U > 0$ is
indistinguishable from a local spin. Charge fluctuations on the site are negligible, while
the local spin interacts antiferromagnetically with spins of conduction electrons via exchange
coupling  $J = V^2/\epsilon_f < 0$. At high temperatures local spins behave as pair-breaking paramagnetic
centers\cite{AG,Anderson}. Below a characteristic temperature, $T_K$, the local moments become screened,
and the Fermi liquid regime sets in \cite{hewson}. For Ce, it corresponds to the $f^1$ configuration.

\noindent
(2) \textbf{The mixed valence regime}, 
\begin{equation}
\label{mvc}
|\epsilon_f| \sim \Gamma.
\end{equation}

In this regime the two configurations, say, $f^0$ and $f^1$ for Ce, are approximately degenerate. The system
is characterized by a time scale for spin fluctuations (charge fluctuations are strong as well).

If the f-level is taken above the chemical potential, and the hybridization, $V$, is weak: 
\begin{equation}
E_0 \simeq \epsilon_f \gg \Gamma, 
\end{equation}
then, the impurity is mostly ``empty'', so that the scattering
is mostly nonmagnetic in character. 
Nevertheless, finite hybridization of electrons with correlated impurity levels introduces an
effective repulsion between conduction electrons with opposite spin, which grows with increased concentration of impurities,
and causes pair weakening. In the
Hartree-Fock approximation such pair weakening caused by 
resonant impurity scattering has been studied  by Kaiser\cite{kaiser1}, Shiba\cite{shiba}, and Schlottmann\cite{schlottmann1}. This
results in a modified exponential decay of $T_c$ with increased impurity concentration\cite{kaiser1}.
The Hartree-Fock approach is only valid for small enough Coulomb repulsion\cite{schlottmann1}, $U/\Gamma \ll 1$. Nevertheless,
it has produced a good description of $ThCe$\cite{maple1} and some lanthanide alloys with large Kondo or
spin fluctuation scales.

 The presence of Kondo effect significantly complicates the treatment of $T_c$ suppression.
The approach\cite{MH} that uses the Nagaoka decoupling only works well
for small values of the Kondo scale $T_K < T_{c0}$. In the Fermi liquid regime, the theory was 
developed by Matsuura, Ichinose, and Nagaoka\cite{MIN}, and by Sakurai\cite{sakurai}. For the Fermi liquid fixed point, 
pair weakening occurs through virtual polarization of the Kondo ground state\cite{MIN,sakurai}. 
This theory is valid when $T_K \gg T_{c0}$. The behavior of magnetic impurities in strongly coupled superconductors has
been studied numerically\cite{jarrell} and analytically\cite{schachinger}, with the result that strong coupling
weakens the effect of Kondo impurities by a factor $1+\lambda$, where $\lambda$ is due to electron-phonon interaction. 
The low-temperature regime has also been
studied in the slave boson $1/N$ formalism\cite{azami}. 
A unified treatment of superconductivity in presence of Anderson impurities in the NCA approximation ($T_c \gg T_K$) has
been done by Bickers and Zwicknagl\cite{bickers}.

Analysis of superconducting properties of alloys in this regime 
has been done perturbatively by  Schlottmann\cite{schlottmann2}:  $T_c$ decreases at first
linearly with concentration, then exponentially. In addition, this regime was studied at zero temperatures\cite{PB}
using the large-degeneracy expansion of Gunnarsson and Sch\"{o}nhammer\cite{GSch}. For the periodic lattice, 
superconductivity in a mixed valence compound was analyzed using the Green function approach\cite{gulacsi}.
Experimentally, superconductivity
in mixed valence regime has been studied in detail for CeRu$_3$Si$_2$\cite{rauchschwalbe},  CeRu$_2$, and CeIr$_3$\cite{hakimi}.

As it was mentioned in introduction, below we study the competition of superconductivity and Kondo effect in concentrated
alloys and Anderson lattices at low temperatures.
The influence of impurities on the superconducting state in heavy fermions has also attracted some experimental interest.
For example, superconductivity in presence of non-magnetic impurities has been studied 
in heavy fermion superconductor CeCu$_2$Si$_2$\cite{steglich1,steglich2}. The peculiar properties of U$_{1-x}$Th$_x$Be$_{13}$
are also well known\cite{fisk1,fisk2}. 
Numerous experiments have been done in other alloys, such as Th$_{1-x}$Ce$_x$, Th$_{1-x}$U$_x$,
Al$_{1-x}$Mn$_x$, La$_{3-x}$Ce$_x$In. 
We refer the reader to Ref.\cite{maple2} for a detailed review of relevant experiments.  However, the theoretical
model developed below leaves scattering effects aside. We apply our results only to systems where the latter
does not play important role, because the f-electrons go into bands 
(i.e., where no sharp decrease of $T_c$ at low concentrations was observed).
For the heavy fermion systems the common assumption since Ref.\cite{Rau} is that SC forms on a heavy Fermi
surface due to electron-electron interactions, and then induces SC on other parts (see, e.g., in Ref. \cite{Lizardo}).
To the contrary, we begin with the phonon mechanism.

\section{Heavy fermion liquid and renormalized bands.}

As usual, we start the consideration of a heavy fermion liquid by writing the periodic Anderson model,
\begin{equation}
H = H_0 + H_V + H_{ef},
\label{theH}
\end{equation}
where
\begin{equation}
H_0 = \sum_{\bm{k} \sigma} \epsilon_k c^{\dagger}_{\bm{k} \sigma} c_{\bm{k} \sigma} + 
\sum_{i \sigma} E_0 f^{\dagger}_{i \sigma} f_{i \sigma} +
\sum_i U f^{\dagger}_{i \uparrow} f^{\dagger}_{i \downarrow} f_{i \downarrow} f_{i \uparrow}. 
\label{fterm}
\end{equation}
Here the creation and annihilation operators for the f-electrons, $f^{\dagger}_{i \sigma}$ and $f_{i \sigma}$,
carry the site index $i$, and there is a Coulomb interaction at each site for the f-electrons. The operators
$c^{\dagger}_{\bm{k} \sigma}$ and $c_{\bm{k} \sigma}$ correspond to delocalized Bloch states.
The hybridization term in the model Hamiltonian, $H_V$, accounts for the $s-d$ hybridization between the  f-electrons and
the Bloch states:
\begin{equation}
 H_V = \sum_{i, \bm{k}, \sigma} \left(
 V_{\bm{k}} e^{i \bm{k} \cdot \bm{R}_i} f^{\dagger}_{i \sigma} c_{\bm{k} \sigma}  +
 V_{\bm{k}}^* e^{-i \bm{k} \cdot \bm{R}_i} c^{\dagger}_{\bm{k} \sigma} f_{i \sigma} \right).
\label{hyb}
\end{equation}
Finally, the third term in the Hamiltonian Eq.(\ref{theH}) corresponds to the attraction,
caused by the electron-phonon interaction, which we consider in the weak coupling limit here:
\begin{equation}
H_{ef} = \frac{\lambda}{2}\,\int \Psi_{\sigma}^{\dagger}(\bm{r})  \Psi_{\sigma'}^{\dagger}(\bm{r}) \Psi_{\sigma'}(\bm{r}) 
\Psi_{\sigma}(\bm{r}) d \bm{r}, 
\label{ef}
\end{equation}
where the $\Psi_{\sigma}(\bm{r})$ and $\Psi_{\sigma}^{\dagger}(\bm{r})$ are operators, which correspond to the
itinerant band:
\begin{eqnarray}
\Psi_{\sigma}(\bm{r}) & = & \frac{1}{(2 \pi)^3}\, \int 
e^{i \bm{k} \cdot \bm{r}} c_{\bm{k} \sigma} d \bm{k}, \\
 \Psi_{\sigma}^{\dagger}(\bm{r}) & = & \frac{1}{(2 \pi)^3}\, \int 
e^{- i \bm{k} \cdot \bm{r}} c_{\bm{k} \sigma}^{\dagger} d \bm{k}. 
\end{eqnarray}

The on-site Coulomb repulsion $U$ is usually very large for f-electron materials, and it will be taken 
infinite below. To account for this, the creation/annihilation operators for $f$-electrons in $H_V$ would have to be taken with
projection operators, which project out doubly occupied $f$-electron states\cite{read}. A convenient way
to rewrite this Hamiltonian, invented by Coleman\cite{col}, Read and Newns\cite{readn}, and Barnes\cite{barnes}, is to introduce
a new slave boson field $b_i^+$, which creates a hole on site $i$, and to rewrite the Anderson Hamiltonian in a
way, which allows a $1/N$ expansion in the number of orbitals. However, in what follows, we resort to a more
phenomenological approach of Edwards\cite{edwards} and Fulde\cite{fulde} (see also Ref.(\cite{hewson})).

Introducing the set of the fermion Green's functions (for imaginary time $\tau$), 
\begin{eqnarray}
G_{cc}^m(\bm{k}, \tau) &\equiv& - \langle T_{\tau} c_{\bm{k} m} (\tau) c^{\dagger}_{\bm{k} m} (0) \rangle, \\
G_{fc}^m(\bm{k}, \tau) &\equiv& - \langle T_{\tau} f_{\bm{k} m} (\tau) c^{\dagger}_{\bm{k} m} (0) \rangle, \\
G_{ff}^m(\bm{k}, \tau) &\equiv& - \langle T_{\tau} f_{\bm{k} m} (\tau) f^{\dagger}_{\bm{k} m} (0) \rangle,
\end{eqnarray}
and transforming to Matsubara frequencies, we find for them the diagrammatic expansion in powers of $U$ and $V_{\bm{k}}$
for the conduction and $f$-electron Green's functions:
\begin{equation}
\label{mattr}
\left[ \begin{array}{cc} 
i \omega_n - \epsilon_f + \mu - \Sigma_{\sigma}(\omega_n, \bm{k}) &  - V_{\bm{k}}\\
- V_{\bm{k}} & i \omega_n - \epsilon_k + \mu
\end{array} \right]
\left[ \begin{array}{cc} 
G_{\bm{k}, \sigma}^{ff}(\omega_n) &  G_{\bm{k}, \sigma}^{cf}(\omega_n)\\
G_{\bm{k}, \sigma}^{fc}(\omega_n) &  G_{\bm{k}, \sigma}^{cc}(\omega_n)
\end{array} \right]
= \hat{I}.
\end{equation}
Assuming that $\Sigma_{\sigma}(\omega_n, \bm{k})$ can be expanded near the 
Fermi surface, $|\bm{k}| = k_F$, and retaining the first order terms:
\begin{equation}
\Sigma(\omega_n, \bm{k}) \simeq \Sigma(0, \bm{k}_F) 
+ (\bm{k} - \bm{k}_F) \cdot \bm{\nabla} \Sigma(0, \bm{k})_{k=k_F} + 
\omega_n \left(\frac{\partial \Sigma(\omega_n, \bm{k}_F)}{\partial \omega_n}\, \right)_{\omega_n = 0},
\label{sigma}
\end{equation}
the Green's function can now be written in the form analogous to the non-interacting
($U=0$) problem:
\begin{equation}
G_{\bm{k}, \sigma}^{cc}(\omega_n) = \frac{1}{i \omega_n - \epsilon_{\bm{k}} + \mu - 
\frac{|\tilde{V}_{\bm{k}}|^2}{i \omega_n - \tilde{\epsilon}_{f\bm{k}}}\, }\,,  
\label{green}
\end{equation}
where 
\begin{eqnarray}
\tilde{\epsilon}_{f \bm{k}} &=& Z (\epsilon_f - \mu + \Sigma^R(0, k_F) + (\bm{k} - \bm{k}_F)\cdot \nabla \Sigma(0, k_F)), 
\label{param}\\
|\tilde{V}_{\bm{k}}|^2 & = & Z |V_{\bm{k}}|^2. \nonumber
\end{eqnarray}
Since $\tilde{\epsilon}_{f \bm{k}}$ is only weakly $\bm{k}$-dependent, we can replace it with a constant, $\epsilon^{eff}_{f}$.
Furthermore, in what follows we also assume that $V_{\bm{k}}$ does not depend on the direction of $\bm{k}$, so that we can
replace $\tilde{V}_{\bm{k}}$ by a constant, $\tilde{V} = \tilde{V}_{k_F}$.
As usual, the quasiparticle residue $Z$ is given by
\begin{equation}
\label{resid}
Z = \frac{1}{1 - \left[\frac{\partial \Sigma(\omega_n, \bm{k}_F)}{\partial \omega_n}\,\right]_{\omega_n = 0}}\,.
\end{equation}
After making use of Eqs (\ref{mattr}) and (\ref{resid}), the Green's functions acquire the form:
\begin{eqnarray}
\nonumber
G_{cc}^m(\bm{k}, \omega_n) &=& \frac{i \omega_n - \epsilon^{eff}_{f}}{(i \omega_n - \epsilon^{eff}_f)(i \omega_n - \xi_{\bm{k}}) - 
\tilde{V}^2}\,, \\
\label{verygreen}
G_{ff}^m(\bm{k}, \omega_n) &=& \frac{i \omega_n - \xi_{\bm{k}}}{(i \omega_n - \epsilon^{eff}_f)(i \omega_n - \xi_{\bm{k}}) - \tilde{V}^2}\,, \\ \nonumber
G_{fc}^m(\bm{k}, \omega_n) &=& \frac{\tilde{V}}{(i \omega_n - \epsilon^{eff}_f)(i \omega_n - \xi_{\bm{k}}) - \tilde{V}^2}\,.
\end{eqnarray}
 From the poles of the Green's functions Eq.(\ref{verygreen}), the renormalized energy spectrum has
the following form:
\begin{equation}
\tilde{\xi}_{k 1,2} = \frac{\epsilon^{eff}_{f} + \xi_k}{2}\, \pm \frac{1}{2}\, 
\sqrt{(\epsilon^{eff}_{f} - \xi_k)^2 + 4 |\tilde{V}|^2},
\label{spec}
\end{equation}
where $\xi_k \equiv \epsilon_k - \mu$. 
In the limit $U \rightarrow \infty$ the effective f-band $\tilde{\epsilon}_{f \bm{k}}$  must lie above
the Fermi surface, so that the total occupation of the f-level, $n_f$, is such that $0<n_f<1$.
For the effective mass of quasiparticles, after expanding $\tilde{\xi}_{k 2}$
in the vicinity of the Fermi surface, one obtains:
\begin{equation}
\frac{m^*}{m} = \frac{|\tilde{V}|^2}{(\epsilon_f^{eff})^2}\, + 1.
\label{mass1}
\end{equation}
In the Kondo limit, $\epsilon_f^{eff} = T_K$ can be thought of as the Kondo temperature.
The conservation of the total number of quasiparticles (Luttinger theorem) leads to the
shift of the chemical potential, given by
\begin{equation}
\tilde{\mu} = \mu +  \frac{|\tilde{V}|^2}{\epsilon_f^{eff}}\,.
\end{equation}
Note, however, that the parameters $\tilde{V}$ and $\epsilon_f^{eff}$ are, in general, temperature-dependent.
Then, their temperature dependence can be studied within a specific model, such as slave boson $1/N$ approach.
Due to the restriction of the slave boson approach to low temperatures, 
our results for phonon-mediated superconductivity are applicable to the case when both 
$\tilde{V}$ and $\epsilon_f^{eff}$ are much greater than $T_c$, so that these parameters could be regarded 
as temperature-independent. Below we merely consider $\tilde{V}$ and $\epsilon_f^{eff} > 0$ as two
free parameters of our theory.

\section{General formula for $T_c$.}

We can now evaluate the superconducting transition temperature, using  the phonon
attraction Hamiltonian Eq.(\ref{ef}), and our new energy spectrum Eq.(\ref{spec}). 
In what follows, we assume that the phonon cutoff $\omega_D$ in  Eq.(\ref{ef}) is much
greater than the Kondo temperature $\epsilon_f^{eff}$, and the effective hybridization $|\tilde{V}|$,
\begin{equation}
\omega_D \gg max\{|\tilde{V}|, \epsilon_f^{eff}\}.
\end{equation}
Following Ref.\cite{AGD}, $T_c$ is obtained by evaluating the Cooper diagram, 
\begin{equation}
T_c \sum_n \int \frac{d \bm{k}}{(2 \pi)^3}\, 
G_{\bm{k}}^{cc}(\omega_n) G_{- \bm{k}}^{cc}(- \omega_n) = \frac{1}{|\lambda|}\,.
\label{Tceq}
\end{equation}
As usual, to get rid of the cutoff dependence arising from the frequency summation, we 
have to introduce a new energy scale $T_{c0}$, for the superconducting temperature in
absence of the f-electrons:
\begin{equation}
T_{c0} = \frac{2 \omega_D \gamma}{\pi}\, e^{- 2 \pi^2/|\lambda| m k_F},
\end{equation}
and rewrite Eq.(\ref{Tceq}) as
\begin{equation}
T_c \sum_{|\omega_n| < \omega_D} \int \frac{d \bm{k}}{(2 \pi)^3}\, 
G_{\bm{k}}^{cc}(\omega_n) G_{- \bm{k}}^{cc}(- \omega_n) = 
T_{c0} \sum_{|\omega_n| < \omega_D} \int \frac{d \bm{k}}{(2 \pi)^3}\, 
G_{\bm{k}}^{0}(\omega_n) G_{- \bm{k}}^{0}(- \omega_n),
\label{Tcc}
\end{equation}
where
\begin{equation}
G_{\bm{k}}^{0}(\omega_n) \equiv \frac{1}{i \omega_n - \xi_k}.
\end{equation}
(Note that we introduced the cutoff $\omega_D$ in the sum over $n$, while the integral over $\xi$ goes
from $ - \infty$ to $+ \infty$.)
After some simple but tedious transformations, we find that the Cooper bubble on the left-hand side
of Eqs(\ref{Tceq}),(\ref{Tcc}) can be written as:
\begin{equation}
\Pi(\omega_n, \bm{k}) \equiv G_{\bm{k}}^{cc}(\omega_n) G_{- \bm{k}}^{cc}(- \omega_n) =
\frac{\omega_n^2 + (\epsilon_f^{eff})^2}{(\omega_n^2 + \tilde{\xi}_{k1}^2)(\omega_n^2 + \tilde{\xi}_{k2}^2)}\,,
\label{simple}
\end{equation}
where $\tilde{\xi}_{k 1,2}$ are given by Eq.(\ref{spec}). Integrating Eq.(\ref{Tcc}) by $\xi_k$, 
we get:
\begin{equation}
T_c \sum_{|\omega_n| < \omega_D} \frac{\pi (\omega_n^2 + (\epsilon_f^{eff})^2}{|\omega_n| (\omega_n^2 + [\epsilon_f^{eff}]^2 + \tilde{V}^2)}\,
= T_{c0}  \sum_{|\omega_n| < \omega_D}  \frac{\pi}{|\omega_n|}\,,
\end{equation} 
which can be rewritten using the definition of the digamma function,
\begin{equation}
\Psi(z) \equiv - \gamma - \sum_{n=1}^{\infty} \left( \frac{1}{z + n - 1}\, - \frac{1}{n}\, \right),
\end{equation}
in the following form:
\begin{equation}
\left[\frac{(\epsilon_f^{eff})^2}{\tilde{V}^2}\, + 1 \right] \ln{\left[\frac{T_c}{T_{c0}}\, \right]} = \Psi\left(\frac{1}{2}\, \right) - 
\frac{1}{2}\, \Psi\left(\frac{1}{2}\, +  i \frac{A}{2 \pi T_c}\, \right)
- \frac{1}{2}\, \Psi\left(\frac{1}{2}\, -  i \frac{A}{2 \pi T_c}\, \right),
\label{fin0}
\end{equation}
where 
\begin{equation}
A = \sqrt{(\epsilon_f^{eff})^2 + \tilde{V}^2} = \epsilon_f^{eff} \sqrt{\frac{m^*}{m}\,}.
\end{equation}
Alternatively, we can use the definition of the effective mass, Eq.(\ref{mass1}), 
to write:
\begin{equation}
\left[\frac{m^*}{m^* - m}\, \right] \ln{\left[\frac{T_c}{T_{c0}}\, \right]} = \Psi\left(\frac{1}{2}\, \right) - 
\frac{1}{2}\, \Psi\left(\frac{1}{2}\, +  i \frac{A}{2 \pi T_c}\, \right)
- \frac{1}{2}\, \Psi\left(\frac{1}{2}\, -  i \frac{A}{2 \pi T_c}\, \right),
\label{fin}
\end{equation}
which is our main result for a stoichiometric compound. 

Let us now analyze this equation in more detail.
First, we note that when $\epsilon_f^{eff} = 0$, Eq.(\ref{fin}) coincides with
the equation for $T_c$ for the paramagnetic pair breaking effect of magnetic field, if we 
replace $A = \mu_B B$. However, unlike for the paramagnetic effect, in general
the factor on the left-hand side of  Eq.(\ref{fin}) is not unity. This means that $T_c \neq 0$
for non-zero $\epsilon_f^{eff} > 0$  for any choice of parameters, even though it could become very small. 

To clarify Eq.(\ref{fin}) more, let us consider the limiting cases. The two obvious cases are
$A \gg T_{c0}$ and $A \ll T_{c0}$.

\noindent
1) $A  \ll T_{c0}$. Then we find:
\begin{equation}
\frac{\Delta T_c}{T_{c0}}\, = - \frac{7 \zeta(3)}{4 \pi^2 T_{c0}^2}\, \tilde{V}^2 =
  - \frac{7 \zeta(3)}{4 \pi^2 T_{c0}^2}\, (\epsilon_f^{eff})^2 \left(\frac{m^*}{m}\, - 1 \right),
\end{equation}
where the coefficient $7 \zeta(3)/4 \pi^2 \simeq 0.21$

\noindent
2) $A  \gg T_{c0}$
Expanding $\Psi$ at high $z$, we get
\begin{equation}
\frac{m}{m^* - m}\, \ln{\left[\frac{T_c}{T_{c0}}\, \right]} = \ln{\left( \frac{\pi T_{c0}}{2 \gamma A}\, \right)}.
\end{equation}

Thus,
\begin{equation}
\frac{T_c}{T_{c0}} = \exp{\left[\frac{m^* - m}{m}\,   \ln{\left(\frac{\pi T_{c0}}{2 \gamma A}\, \right)} \right]}
=  \left(\frac{\pi T_{c0}}{2 \gamma A}\, \right)^{(m^*/m) - 1}.
\end{equation}

\begin{figure}
\includegraphics[width=5in]{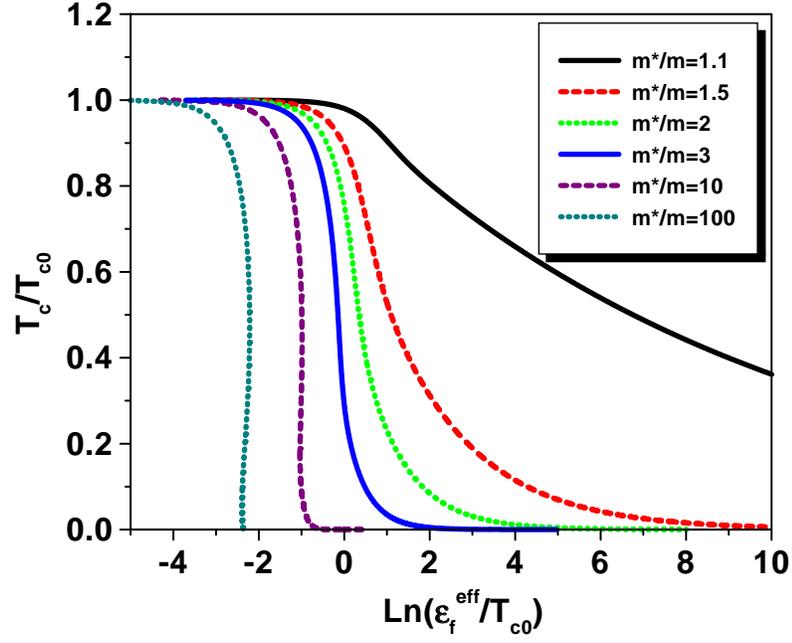}
\caption{Superconducting transition temperature as a function of parameters $m^*/m$ and $\epsilon_f^{eff}/T_{c0}$.}
\label{tcfig}
\end{figure}

Finally, we note that Eq.(\ref{fin}) can be written in terms of dimensionless quantities, 
\begin{equation}
t \equiv \frac{T_c}{T_{c0}}\,, \ \ \ \tilde{A} \equiv \frac{A}{T_{c0}}\,.
\end{equation}
Indeed, we can write it now as:
\begin{equation}
\left[\frac{m^*}{m^* - m}\, \right] \ln{t} = \Psi\left(\frac{1}{2}\, \right) - 
\frac{1}{2}\, \Psi\left(\frac{1}{2}\, +  i \frac{\tilde{A}}{2 \pi t}\, \right)
- \frac{1}{2}\, \Psi\left(\frac{1}{2}\, -  i \frac{\tilde{A}}{2 \pi t}\, \right).
\label{fin1}
\end{equation}
We plot Eq.(\ref{fin1}) as a function of two parameters, $m^*/m$ and $\epsilon_f^{eff}/T_{c0}$,
in Fig \ref{tcfig}. 

\begin{figure}
\includegraphics[width=5in]{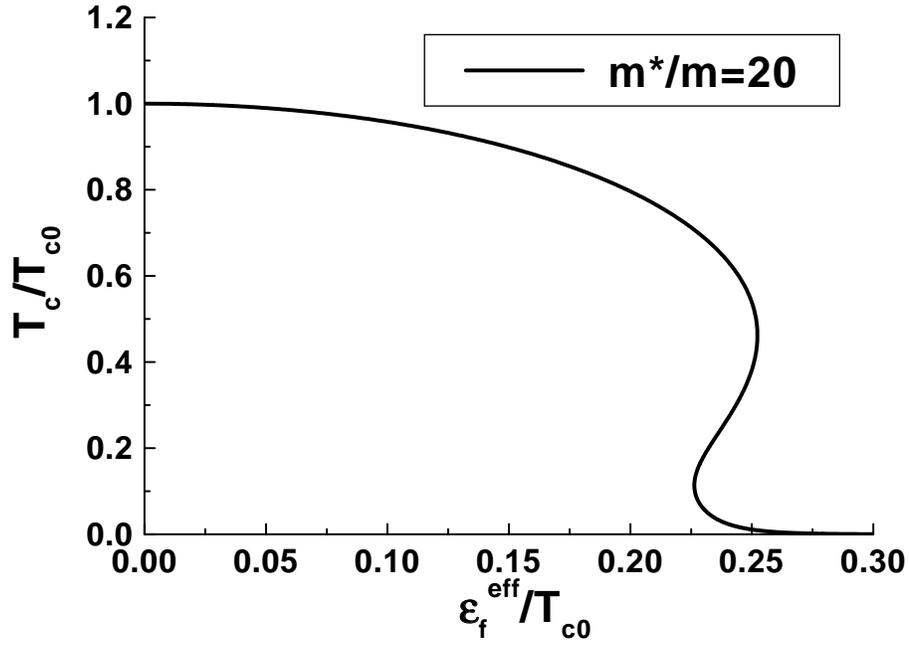}
\caption{Re-entrant behavior of superconducting transition temperature as a function of parameter $\epsilon_f^{eff}/T_{c0}$.}
\label{tcfig2}
\end{figure}

Interesting enough, the dependence of $T_c$ on parameters may have a characteristic
S-shape re-entrance form, as shown more explicitly in Fig \ref{tcfig2}. This, and
the similarity of Eqs (\ref{fin0}),(\ref{fin}) to the field dependence for paramagnetic pair breaking raises two questions:

\vglue 0.5 cm

\noindent
a) Is the superconducting transition always second order?

\noindent
b) Is the homogeneous state the most stable state, or could a superstructure
similar to Larkin-Ovchinnikov-Fulde-Ferrel (LOFF)\cite{FF,LO} state appear?

\vglue 0.5 cm

We have investigated both questions, and found that the superconducting transition
is always of the second order, and that the inhomogeneous state is always energetically 
unfavorable. As to the S-shape dependence in Figs \ref{tcfig}, \ref{tcfig2}, one sees that it arises
at rather specific conditions: $\epsilon_f^{eff}$ must be less than $T_{c0}$ (usually, of the order of a few K).
according to Eq.(\ref{mass1}), $\tilde{V}$ is also rather weak. For a single level center it 
corresponds to Eq. (\ref{mvc}) with $\Gamma \ll 1$, i.e., an almost localized level near
the chemical potential that can be thermally populated. We are not aware of a material that
would satisfy these criteria.

\section{Superconductivity in dense alloys.}

Since experiments often deal with alloys, let us consider in
detail the dependence of the critical temperature on concentration in the case
when $\epsilon_f^{eff}$ and $m^*/m$ are fixed. This
applies, for example, to dense mixed valence alloys in the phenomenological
description of Eqs (\ref{green})-(\ref{spec}).
For the concentration of f-elements per unit cell $x$ ($x < 1$), after repeating
the above calculations, we find for $T_c$:
\begin{equation}
\left[\frac{(\epsilon_f^{eff})^2}{x \tilde{V}^2}\, + 1 \right] \ln{\left[\frac{T_c}{T_{c0}}\, \right]} = 
\Psi\left(\frac{1}{2}\, \right) - 
\frac{1}{2}\, \Psi\left(\frac{1}{2}\, +  i \frac{A(x)}{2 \pi T_c}\, \right)
- \frac{1}{2}\, \Psi\left(\frac{1}{2}\, -  i \frac{A(x)}{2 \pi T_c}\, \right),
\label{tcalloy}
\end{equation}
where 
\begin{equation}
A(x) = \sqrt{(\epsilon_f^{eff})^2 + x \tilde{V}^2}.
\end{equation}
In the derivation Eqs(\ref{mattr})-(\ref{green}), 
the level $\epsilon_f^{eff}$ and the effective hybridization $\tilde{V}$ correspond to  
the Anderson model for a single impurity, and do not change with concentration. However, we see 
that the concentration of f-elements enters explicitly in our new equation, with $x$ as the probability
to find the $f$-level at a given site. This equation is, in fact, the
same as the solution for the lattice problem, with $\tilde{\tilde{V}} \equiv \sqrt{x} \tilde{V}$  used
as the effective hybridization. In other words, in alloys effective hybridization is ``tuned up'' by the
concentration of $f$-levels. 

At small, but finite $\epsilon_f^{eff}$, i.e., in the Kondo regime, one is able, like in Fig. \ref{tcfig2},
to obtain the S-shape of $T_c$ - this time, however, as a function of concentration, i.e., the regime
with three transition temperatures, which was the main result in Ref.\cite{MH}. Note that the approach
in Ref.\cite{MH}, strictly speaking, was not rigorously founded at concentrations for the 
re-entrant superconductivity regime. As for our Eq.(\ref{tcalloy}), although it also predicts an initial
decrease of $T_c$ at small x (and small $\epsilon_f^{eff}$), as in Eq.(\ref{fin}), it is also deficient here,
because the electron \textit{scattering} on $f$-centers is not included in the framework of our
derivation. 

In what follows, we only discuss the case $T_c \ll \epsilon_f^{eff}$, since it applies to most experiments
on dense alloys. Then we don't expect our parameters, $\epsilon_f^{eff}$ and $\tilde{V}$, to be temperature-dependent.
Since $T_{c0} \ll \epsilon_f^{eff}$, we are always in the limit $A \gg T_{c0}$. Thus, the decay of $T_c$ with
concentration is exponential. Let us introduce, for convenience, two new dimensionless parameters, $k$ 
and $y$:
\begin{eqnarray}
k & = & \frac{2 \gamma \epsilon_f^{eff}}{\pi T_{c0}}\,, \nonumber \\
y & = & \frac{m^*(x=1)}{m}\, = \frac{\tilde{V}^2}{(\epsilon_f^{eff})^2}+1\,. 
\end{eqnarray}
Then we can write $T_c$ as a function of doping as:
\begin{equation}
\frac{T_c}{T_{c0}}\, = \exp{\left[- x(y-1) \ln{\left(k \sqrt{1 + x(y-1)} \right)} \right]}.
\end{equation}

\begin{figure}
\includegraphics[width=5in]{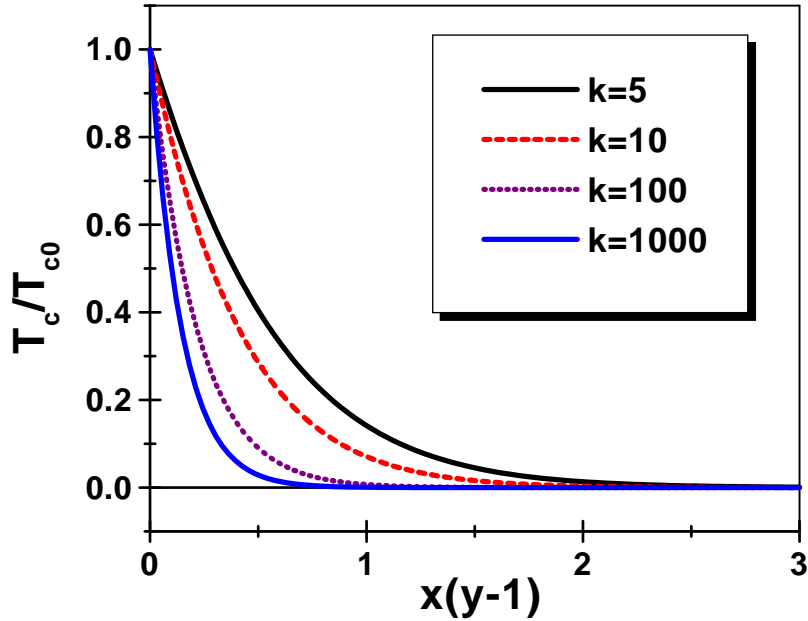}
\caption{Superconducting transition temperature as a function of doping $x$ in alloys.}
\label{tcm}
\end{figure}

The doping dependence of $T_c$ for various choices of parameter $k$
is shown in Fig. \ref{tcm}.
We can see that characteristic concentration where $T_c$ decays substantially depends
mainly on the ratio of $\epsilon_f^{eff}$ and $\tilde{V}$. Thus, for Kondo impurities, where 
$\epsilon_f^{eff} \ll \tilde{V}$, $T_c$ is rapidly suppressed, although remains non-zero at
$x=1$. On the other
hand, for mixed valence impurities $ \epsilon_f^{eff} \sim \tilde{V}$, so 
$T_c$ is suppressed exponentially, but not so steep.

\section{Discussion of the experimental data.}

As it would follow from our results, if a phonon (s-wave) superconductivity were present
for a compound without an element with f-electrons (e.g., La or Lu), one should expect that
its stoichiometric homolog with the f-electrons being present, will have a small, but finite,
$T_c \neq 0$. This prediction is, of course, not fully conclusive, for already at dilute solution
of paramagnetic centers the pair-breaking scattering initially may take $T_c$ practically to zero,
while appearing again in the periodic (dense) limit in absence of scattering.
Our approach, however, is to circumvent the scattering regime, and in its framework one may estimate
the phonon superconductivity $T_c$ at the other end through Kondo, or, more generally, the
Anderson-lattice parameters of the latter. Certainly, superconductivity in a heavy fermion compound may be due
to a completely different mechanism, but the analysis along the line proposed above, nevertheless, seems
to be instructive. Thus, for instance, LaCu$_2$Si$_2$ is not a superconductor, while CeCu$_2$Si$_2$\cite{steglich}
is. Similarly, there are no SC homologs without $f$-electrons known for UBe$_{13}$. 
This gives an additional and unambiguous argument in favor of an unconventional superconductivity
mechanism for these HF materials.

Unfortunately, such decisive experimental data are rather scarce and are available mostly with the Ce-based
materials. In the literature the latter are often subdivided into ``strongly mixed valent'' (CeRu$_2$; CeCo$_2$)
and ``weakly mixed valent'' (CeSn$_3$, CePd$_3$, CeBe$_{13}$) compounds
(see, e.g., in Ref.\cite{rauchschwalbe}). Among the second group no superconductors have been found, and they
are usually considered to be closer to localized moments, or Kondo, materials. Among the first group we have
pairs:
CeRu$_2$ ($T_c \simeq 6K$) and LaRu$_2$($T_c \simeq 4K$), LaPd$_2$Ge$_2$ ($T_c \simeq 1.1K$)\cite{Hull}
and CePd$_2$Ge$_2$ ($T_c \simeq 0.07K$, $P = 14.6 GPa$)\cite{Wilhelm}, LaNi$_2$Ge$_2$ ($T_c \simeq 0.8K$)\cite{Maezawa}
and CeNi$_2$Ge$_2$ ($T_c \simeq 0.2K$)\cite{Gegenwart}. No superconductivity has ever been found for Yb compounds\cite{kishi}.
Common view (e.g., in Ref.\cite{hakimi}) is that in mixed valence compounds f-electrons merely hybridize with
one of conduction bands, thus supplying additional electrons into that band. 

\begin{figure}
\includegraphics[width=5in]{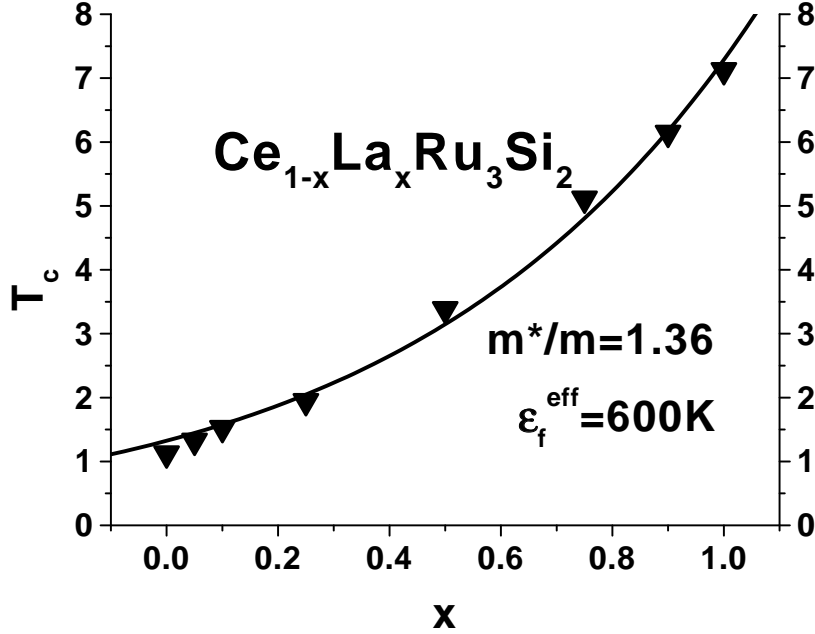}
\caption{Superconducting transition temperature as a function of doping $x$ in Ce$_{1-x}$La$_x$Ru$_3$Si$_2$,
from Ref.\cite{rau17} vs our theoretical fit.}
\label{expe}
\end{figure}

We consider in more detail CeRu$_3$Si$_2$ and its homolog LaRu$_3$Si$_2$ with its Ru-derived 4d-band, where
the data for the continuous alloying are available\cite{rau17}. In Fig.\ref{expe} we show the fit for $T_c$ in
the Ce$_{1-x}$La$_x$Ru$_3$Si$_2$ in the whole concentration range $x$, making use of our Eq.(\ref{tcalloy}).
One sees that the experimental behavior is well reproduced at $\epsilon_f^{eff} = 600K$ and $m^*/m = 1.36$.
Small mass ratio is well explained in terms of large band masses for $Ru$ d-electrons\cite{kishi,kishi1}.
The material provides an example where scattering at small concentrations (on the both sides) plays no
significant role on $T_c$. That is also in favor of common SC mechanisms (s-wave).

\section{Summary.}

In conclusion, we have considered phonon (s-wave) superconductivity in the Anderson lattice. Our result indicates
that the Kondo effect competes with Cooper pairing, and leads to a dramatic reduction of $T_c$.
We found that $T_c$ may have a re-entrant behavior at somewhat exotic choice of parameters, and that such behavior is
stable. We note that our result is similar, but different from the results of Ref.\cite{MH} on Kondo alloys,
since it does not map on the scattering from paramagnetic impurities \cite{AG}. There are no scattering processes 
in Kondo lattices. Instead, the behavior is coherent. Our results correspond to the 
low-temperature coherent Fermi-liquid fixed point, which
does not map readily on the high-temperature case of free paramagnetic spins. We caution that
we consider $\epsilon_f^{eff}$ and $\tilde{V}$ as 
phenomenological parameters.
 
In the case of mixed valence, and lower effective mass $m^* \sim m_e$, we find that $T_c$ is substantially less suppressed.
Therefore, $T_c$ as a function of concentration does not exhibit re-entrant behavior. 

On a more qualitative level we would like to add a comment that physics may be different in real systems due
to the presence of  more than just a single band. It is quite possible that, while one band experiences large
mass renormalization via the Anderson hybridization with the f-electrons, the second one does not. At the
same time,  superconductivity there that is due to a common s-wave mechanism may induce a considerable 
superconducting gap on the ``heavy fermion'' band. For instance, in Ref.\cite{ABG} it was shown theoretically
that such a multi-band view could even lead to appearance of a non-trivial superconducting order parameter for
specific electron bands spectra. Thermodynamics of the superconducting transition reflects
the ``heaviness'' of that first band, while $T_c$ itself continues to be due to the mechanisms for the
light band. (For results on thermodynamics of multi-band superconductors, see Ref. \cite{2band}; these results
were recently re-derived by Zhitomirsky and Dao\cite{Zhit}).
An example of such a homolog pair may be given by PrOs$_4$Sb$_{12}$ ($T_c \simeq 1.85K$) and
LaOs$_4$Sb$_{12}$ ($T_c \simeq 1 K$) \cite{maple3}.

One of the authors (LPG) acknowledges the partial support from the Alexandr Von Humboldt Foundation
and the hospitality of the Max Plank Institute for Chemical Physics, where this work has
been initiated. We are grateful for numerous discussions and helpful information to Z. Fisk, P. Gegenwart,
C. Geibel, Yu. Grin, M. B. Maple, J. Mydosh, F. Steglich, and H. Wilhelm.

This work was supported (VB) by TAML at the University of Tennessee and (LPG) by NHFML through
the NSF Cooperative agreement No. DMR-008473 and the State of Florida.

\end{document}